\documentclass[epj]{svjour}

\usepackage{epsfig,graphicx}

\begin{document}

\title{ Nucleon elastic scattering in quark-diquark representation with springy Pomeron} 
\author{ 
V.M.~Grichine\thanks{ 
e-mail:  Vladimir.Grichine@cern.ch}
}
\institute{
Lebedev Physical Institute, Moscow, Russia 
}

\date{Received: date / Revised version: date}

\abstract{
A model for elastic scattering of nucleons (and anti-nucleons)  based on the quark-diquark 
representation of the nucleon with springy Pomeron, providing increased real part of the 
scattering amplitude, is developed. The model predictions are compared with experimental 
data for the  differential elastic cross-sections of nucleons in the energy range from 
few GeV up to 7 TeV using available databases. 
}

\maketitle

\section{Introduction}
\label{intro}

The quark-diquark (qQ-) model for description of the proton-proton elastic scattering proposed in~\cite{vat79} 
was recently updated and compared with experimental data in~\cite{qQ2013}. Fig. 1 shows the differential 
cross-section of proton-proton elastic scattering corresponding to the total energy in the center of mass system, 
$\sqrt{s}$=7~TeV. The curve corresponds to the qQ-model~\cite{qQ2013} with the standard Pomeron parametrization:
\begin{equation}
\label{sP}
\exp\left\{\tilde{\alpha}\left[\ln\frac{s}{s_o}-\frac{i\pi}{2}\right]\right\},
\end{equation}
where the Pomeron trajectory slope $\tilde{\alpha}$=0.15~GeV$^{-2}$ and the 
term $-i\pi/2$ defines the scattering amplitude real part of the qQ-model~\cite{qQ2013} 
($s_o$=1~GeV$^2$; we use units: $\hbar=c=1$). 
It is seen that the curve overestimates the dip value, and the reason is that the scattering amplitude 
real part value is not big enough. 

The picture here is similar to the elastic scattering of hadrons on nuclei. Fig. 2 shows 
the differential hadron elastic scattering cross-sections of protons with the energy 
1 GeV on lead versus the polar scattering angle~\cite{diffuse}. The pure diffuse diffraction model 
proposed in~\cite{diffuse} does not describe the minimum regions of the differential cross-section. 
If we add, however, the Coulomb amplitude, which increases the scattering amplitude real part, 
the model becomes to be more close to the experimental data. The Coulomb amplitude can increase the 
scattering amplitude real part for heavy nuclei with high atomic number only. In the case of nucleons 
one should modify directly the hadronic amplitude. Therefore one 
can assume, that if we increase the imaginary part of the Pomeron parametrization (\ref{sP}), i.e. apply:
\begin{equation}
\label{mP}
\exp\left\{\tilde{\alpha}\left[\ln\frac{s}{s_o}-\alpha_p\frac{i\pi}{2}\right]\right\},
\end{equation}
with the empirical parameter, $|\alpha_p|>1$, which can be named as the Pomeron elasticity, 
we can improve the description of the hadron-hadron elastic scattering. This modification is proposed 
in the current paper together with an extension of the qQ-model to the case of the elastic 
scattering of different nucleons (and even hadrons).

We discuss below the main features of the extended qQ-model suitable for numerical 
calculations of elastic scattering of different nucleons and provide comparisons with the experimental 
data for the $pp$, $np$, and $\bar{p}p$ elastic differential cross-section in the energy 
range from few GeV up to 7~TeV in the center of mass system.


\begin{figure}
\includegraphics[height=2.8in,width=3.5in]{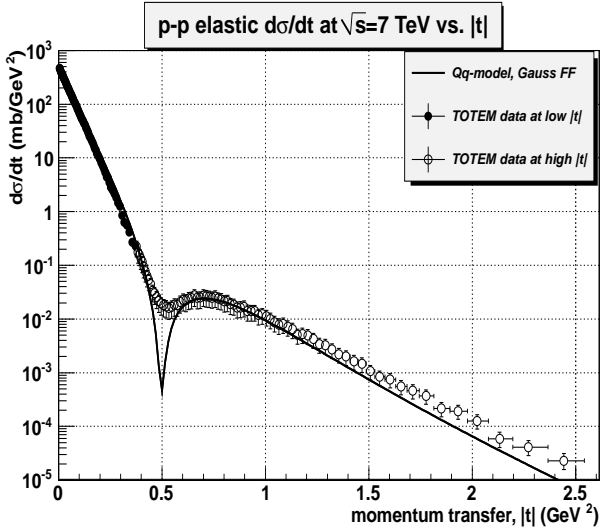}
\caption{The proton-proton differential elastic cross-section 
versus $|t|$ at $\sqrt{s}=$7~TeV. The curve is the prediction of quark-diquark model~\cite{qQ2013}. 
The open and closed circles are the LHC TOTEM experimental data 
from~\cite{totem1}. }
\label{old7tev}
\end{figure}

\begin{figure}
\centering \includegraphics[width=3.5in,height=2.8in]{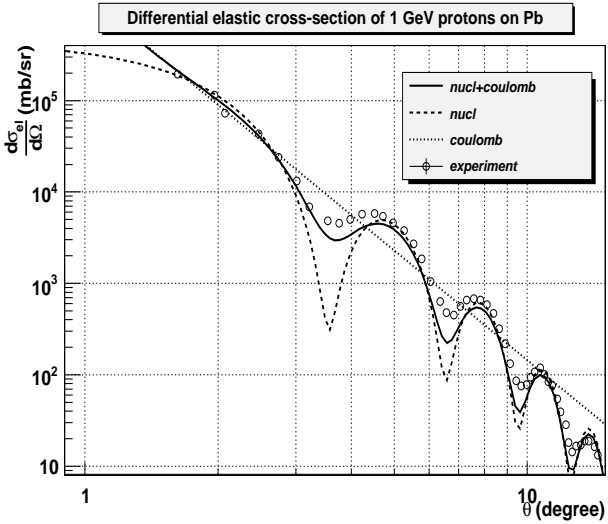}
\caption{ 
The differential hadron elastic scattering cross-sections of protons with the energy 
1 GeV on lead versus the polar scattering angle~\cite{diffuse}. The curves show the pure nuclear 
(dash-dash) and with the Coulomb correction (solid) models. The dot-dot line 
corresponds to the pure electromagnetic Coulomb scattering. The open points are 
experimental data.}
\label{pPb1GeV}
\end{figure}


\section{Quark-diquark model for nucleon-nucleon elastic scattering}

The nucleon-nucleon differential elastic cross-section, $d\sigma_{el}/dt$, can be expressed in terms of the scattering 
amplitude $F(s,t)$:
\begin{equation}
\label{dsdt}
\frac{d\sigma_{el}}{dt}=\frac{\pi}{p^2}|F(s,t)|^2, 
\end{equation}
where $p$ is the nucleon momentum in the center of mass system, 
and $t$ is the four-momentum transfer squared. 

We consider here an extension of the model~\cite{qQ2013} describing the elastic scattering of two 
different nucleons. The first nucleon consists of quark (index 1), diquark (2), and the second 
one consists of quark (3) and diquark (4). The model~\cite{qQ2013} limits the consideration of 
the scattering amplitude by contributions from one- 
and two-Pomeron exchanges between quark-quark (1-3), diquark-diquark (2-4) and two
quark-diquark (1-4,~2-3). In this approximation  $F(s,t)$ can be expressed as:
\begin{equation}
\label{f}
F(s,t)=F_1(s,t)-F_2(s,t)-F_3(s,t),
\end{equation}
where $F_1(s,t)$ is the scattering amplitude with one-Pomeron exchange, while $F_2(s,t)$ 
corresponds to two-Pomeron exchanges between the nucleon constituents, quark and diquark, 
and $F_3(s,t)$ corresponds to two-Pomeron exchanges between the quark (or diquark) of 
one nucleon and the  quark and the diquark of another nucleon at the same time. The 
amplitude $F_1(s,t)$ reads:
\begin{equation}
\label{f1}
F_1(s,t)=\frac{ip\sigma_{tot}(s)}{4\pi}\left[f_{13}+f_{14}+f_{23}+f_{24}\right],
\end{equation}
\[
f_{13}=B_{13}\exp[(\xi_{13}+\beta^2\lambda+\delta^2\eta)t], 
\]
\[
f_{14}=B_{14}\exp[(\xi_{14}+\beta^2\lambda+\gamma^2\eta)t], 
\]
\[
f_{23}=B_{23}\exp[(\xi_{23}+\alpha^2\lambda+\delta^2\eta)t], 
\]
\[
f_{24}=B_{24}\exp[(\xi_{24}+\alpha^2\lambda+\gamma^2\eta)t], 
\]
where $\sigma_{tot}(s)$ is the total  nucleon-nucleon cross-section. The 
coefficients $B_{jk}$, parametrize the quark-quark, diquark-diquark and quark-diquark
cross-sections:
\[
\sigma_{13}=B_{13}\sigma_{tot}(s), \quad\sigma_{24}=B_{24}\sigma_{tot}(s),
\]
\[
\sigma_{23}=B_{23}\sigma_{tot}(s), \quad\sigma_{14}=B_{14}\sigma_{tot}(s).
\]
The model assumes the quark-diquark cross-section, $\sigma_{23}=\sigma_{14}=\sqrt{\sigma_{13}\sigma_{24}}$. 
The coefficients $\alpha=\gamma=1/3$, and $\beta=\delta=2/3$ correspond to the relative masses of 
quark and diquark in the nucleons. The parameters $\lambda$ and $\eta$ are equal to the the first 
and second nucleon radius squared divided by four, respectively.

The coefficients $\xi_{jk}$, ($j,k=1,2$) are derived taking into account the Gauss distribution 
of quark and diquark in nucleon together with the Pomeron parametrization discussed above in 
relation~(\ref{mP}). They read:
\begin{equation}
\label{Ajk}
\xi_{jk}=\frac{r^2_j+r^2_k}{16}+\tilde{\alpha}\left[\ln\frac{s}{s_o}-\alpha_p\frac{i\pi}{2}\right].
\end{equation}
Here  $r_j$, $r_k$ are the quark or diquark radii.  The quark and diquark radii 
$r_{1}$ ($r_{3}$) and $r_{2}$ ($r_{4}$) were found by the fitting of experimental data to be 
$0.173$ and $0.316$ of the corresponding nucleon radius, respectively. 

The amplitudes $F_2(s,t)$ and $F_3(s,t)$ are:
\begin{equation}
\label{f2}
F_2(s,t)=\frac{ip\sigma^2_{tot}(s)}{16\pi^2}[f_{13,24}+f_{14,23}],
\end{equation}
\[
f_{13,24}=\frac{B_{13}B_{24}}{\xi_{13}+\xi_{24}+\lambda+\eta}
\exp\left\{
\left[\xi_{24}+\alpha^2\lambda+\gamma^2\eta-\right.\right.
\]
\[
-\left.\left.\displaystyle\frac{(\xi_{24}+\alpha\lambda+\gamma\eta)^2}{\xi_{13}+\xi_{24}+\lambda+\eta}
\right]t\right\},
\]
\[
f_{14,23}=\frac{B_{14}B_{23}}{\xi_{14}+\xi_{23}+\lambda+\eta}
\exp\left\{
\left[\xi_{23}+\alpha^2\lambda+\delta^2\eta-\right.\right.
\]
\[
-\left.\left.\displaystyle\frac{(\xi_{23}+\alpha\lambda+\delta\eta)^2}{\xi_{14}+\xi_{23}+\lambda+\eta}
\right]t\right\},
\]
and:
\begin{equation}
\label{f3}
F_3(s,t)=\frac{ip\sigma^2_{tot}(s)}{32\pi^2}[f_{13,14}+f_{23,24}+f_{13,23}+f_{14,24}],
\end{equation}
\[
f_{13,14}=\frac{B_{13}B_{14}}{\xi_{13}+\xi_{14}+\eta}
\exp\left\{
\left[\xi_{14}+\beta^2\lambda+\gamma^2\eta-\right.\right.
\]
\[
-\left.\left.\displaystyle\frac{(\xi_{14}+\gamma\eta)^2}{\xi_{13}+\xi_{14}+\eta}
\right]t\right\},
\]
\[
f_{23,24}=\frac{B_{23}B_{24}}{\xi_{23}+\xi_{24}+\eta}
\exp\left\{
\left[\xi_{24}+\alpha^2\lambda+\gamma^2\eta-\right.\right.
\]
\[
-\left.\left.\displaystyle\frac{(\xi_{24}+\gamma\eta)^2}{\xi_{23}+\xi_{24}+\eta}
\right]t\right\},
\]
\[
f_{13,23}=\frac{B_{13}B_{23}}{\xi_{13}+\xi_{23}+\lambda}
\exp\left\{
\left[\xi_{23}+\alpha^2\lambda+\delta^2\eta-\right.\right.
\]
\[
-\left.\left.\displaystyle\frac{(\xi_{23}+\alpha\lambda)^2}{\xi_{13}+\xi_{23}+\lambda}
\right]t\right\},
\]
\[
f_{14,24}=\frac{B_{14}B_{24}}{\xi_{14}+\xi_{24}+\lambda}
\exp\left\{
\left[\xi_{24}+\alpha^2\lambda+\beta^2\eta-\right.\right.
\]
\[
-\left.\left.\displaystyle\frac{(\xi_{24}+\alpha\lambda)^2}{\xi_{14}+\xi_{24}+\lambda}
\right]t\right\},
\]
respectively. 

To simplify the numerical calculations, one can assume that all nucleons involved to the 
elastic scattering have the same radius and the quark-quark and diquark-diquark cross-sections do not 
depend on the what quarks ($u$ or $d$, as well as $\bar{u}$ or $\bar{d}$) interact. Then the quark-quark 
cross-section, $\sigma_{13}$, the nucleon radius and the Pomeron elasticity, $\alpha_p$, are the free 
parameters defining (together with $\sigma_{tot}(s)$) the $s$-dependence of the $d\sigma_{el}/dt$. 
The  diquark-diquark cross-section, $\sigma_{24}$, and the parameter  $B_{24}$ are derived from the optical theorem, 
as it was discussed in~\cite{qQ2013}. The s-dependencies of the nucleon radius and the quark-quark cross-section are 
essentially the same as it was shown in~\cite{qQ2013}. The averaged s-dependence of the Pomeron elasticity, $\alpha_p$ 
is shown in fig. 3.


\begin{figure}
\includegraphics[height=2.8in,width=3.5in]{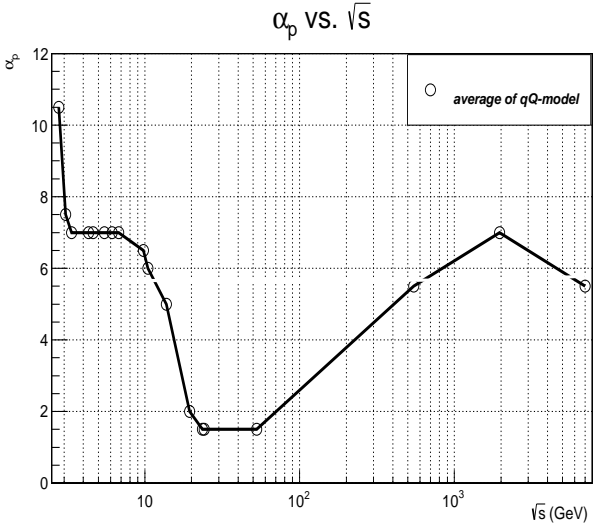}
\caption{ The dependence of $\alpha_p$  versus the $\sqrt{s}$.}
\label{Rpvs}
\end{figure}
 

\section{Comparison with experimental data}

\begin{figure}
\includegraphics[height=2.8in,width=3.5in]{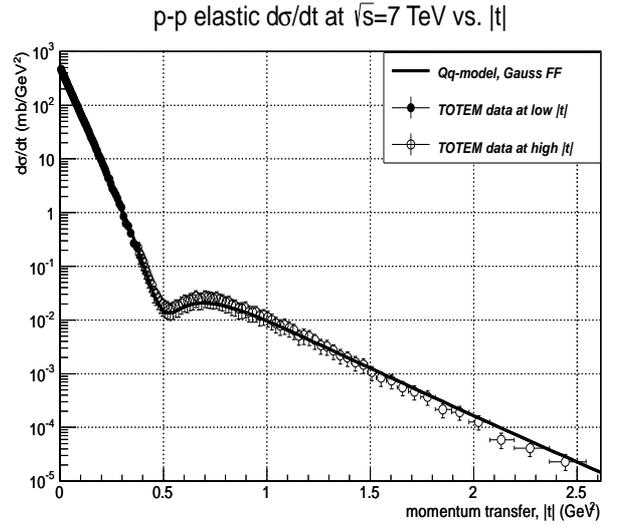}
\caption{The proton-proton differential elastic cross-section 
versus $|t|$ at $\sqrt{s}=$7~TeV. The curve is the prediction of quark-diquark model 
with springy Pomeron. The open and closed circles are the LHC TOTEM experimental data 
from~\cite{totem1}. }
\label{7tev}
\end{figure}

\begin{figure}
\includegraphics[height=2.8in,width=3.5in]{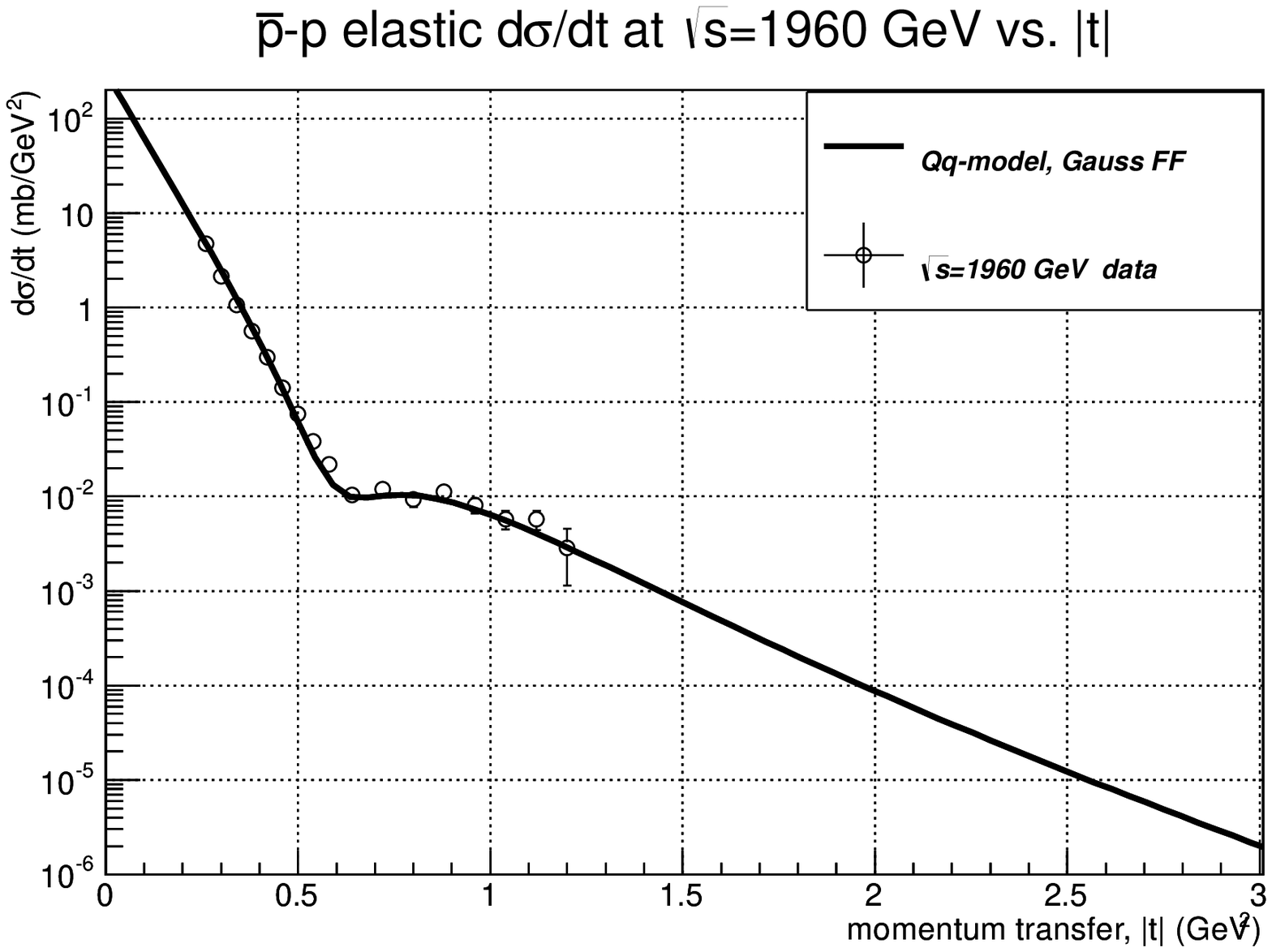}
\caption{The antiproton-proton differential elastic cross-section 
versus $|t|$ at $\sqrt{s}=$1960~GeV. The curve is the prediction of our model. 
The open circles are the experimental data~\cite{pbp1960}. }
\label{1960gev}
\end{figure}

\begin{figure}
\includegraphics[height=2.8in,width=3.5in]{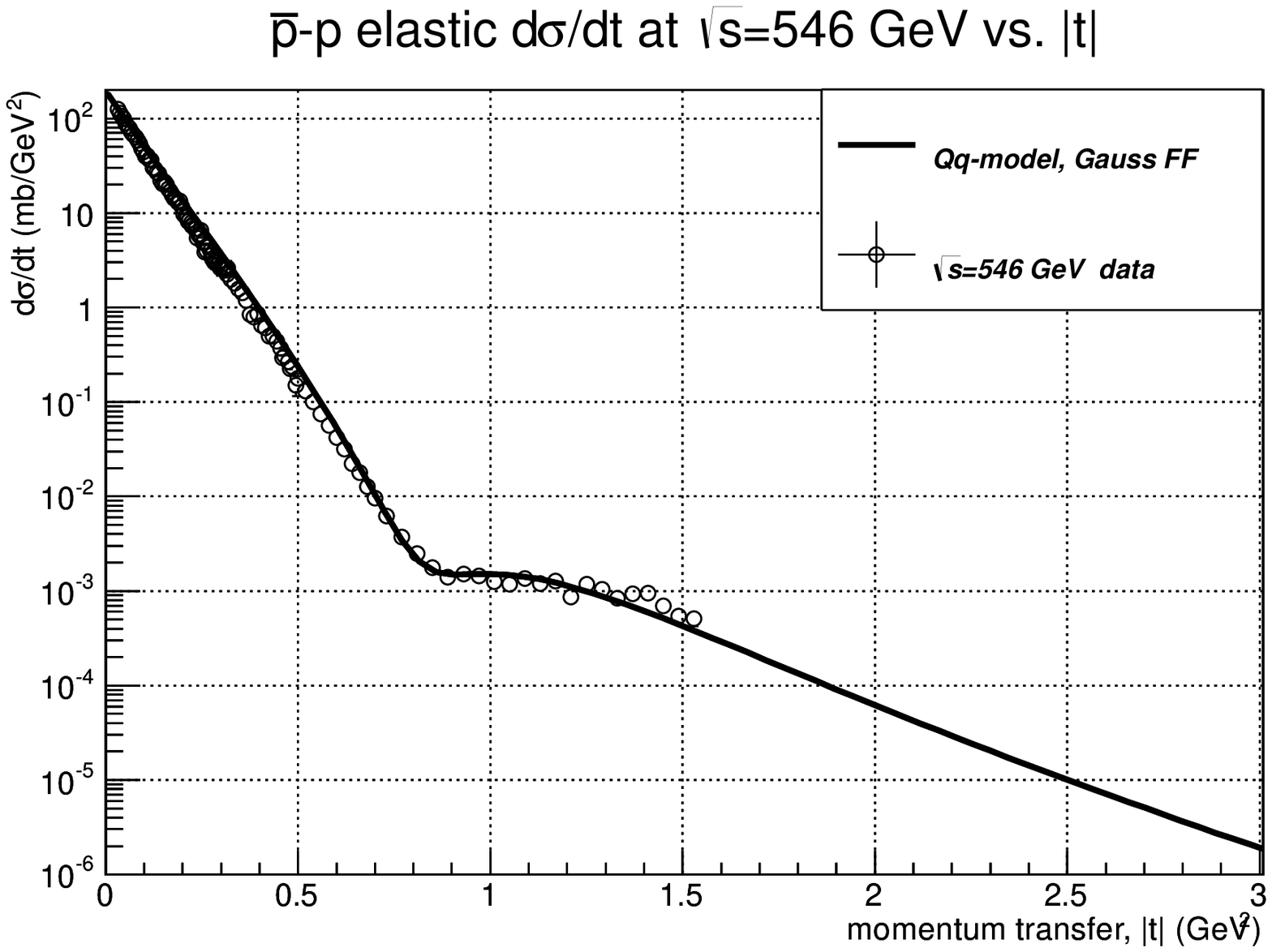}
\caption{The antiproton-proton differential elastic cross-section 
versus $|t|$ at $\sqrt{s}=$546~GeV. The curve is the prediction of our model. 
The open circles are the experimental data~\cite{pbp546}. }
\label{546gev}
\end{figure}

Fig.~\ref{7tev}, \ref{1960gev}, and \ref{546gev} show the proton-proton 
differential elastic cross section at $\sqrt{s}=$7~TeV versus $|t|$ and the antiproton-proton differential elastic 
cross-sections at, $\sqrt{s}=$1960~GeV and  $\sqrt{s}=$546~GeV, respectively.  The curves are the 
predictions of our model. We see that the proposed $qQ$-model with springy Pomeron describes reasonably the 
differential elastic cross sections of the antiproton-proton and proton-proton scattering
in the TeV-region of energy. 

\begin{figure}
\includegraphics[height=2.8in,width=3.5in]{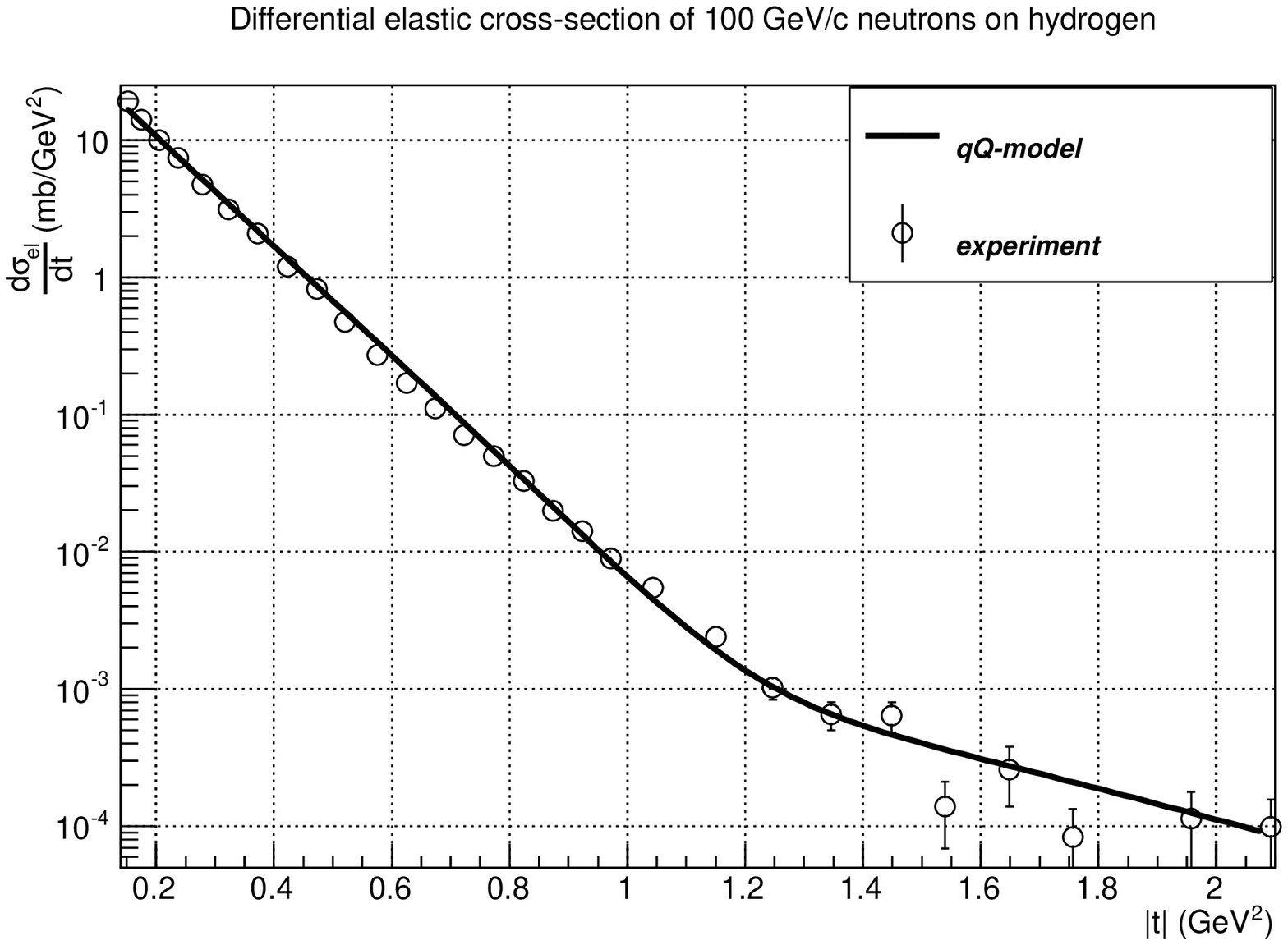}
\caption{The neutron-proton differential elastic cross-section 
versus $|t|$ at the neutron momentum in the laboratory system 100~GeV/c. The curve is the prediction of our model. 
The open circles are the experimental data~\cite{np100gevc}. }
\label{np100}
\end{figure}

\begin{figure}
\includegraphics[height=2.8in,width=3.5in]{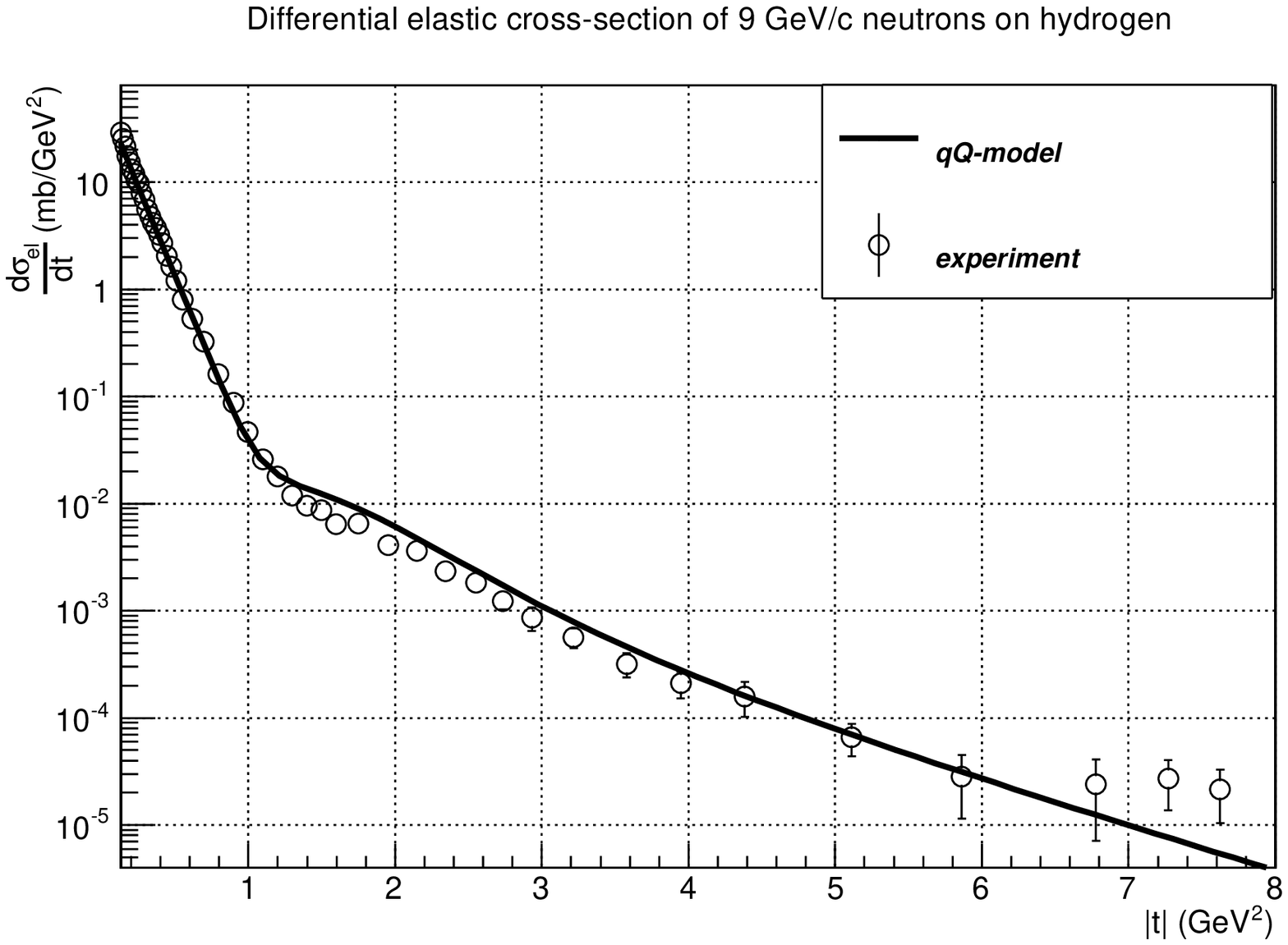}
\caption{The neutron-proton differential elastic cross-section 
versus $|t|$ at the neutron momentum in the laboratory system 9~GeV/c. The curve is the prediction of our model. 
The open circles are the experimental data~\cite{np9gevc}. }
\label{np9}
\end{figure}

Fig.~\ref{np100} and~\ref{np9} show the neutron-proton differential elastic cross sections at the neutron 
momentum in the laboratory system 100 and 9 GeV/c, respectively. Fig. \ref{pp3} the proton-proton 
differential elastic cross-section versus $|t|$ at the proton momentum in the laboratory system 3~GeV/c. 
One can see the satisfactory agreement of the qQ-model with experimental data in the GeV-range of energy.


\begin{figure}
\includegraphics[height=2.8in,width=3.5in]{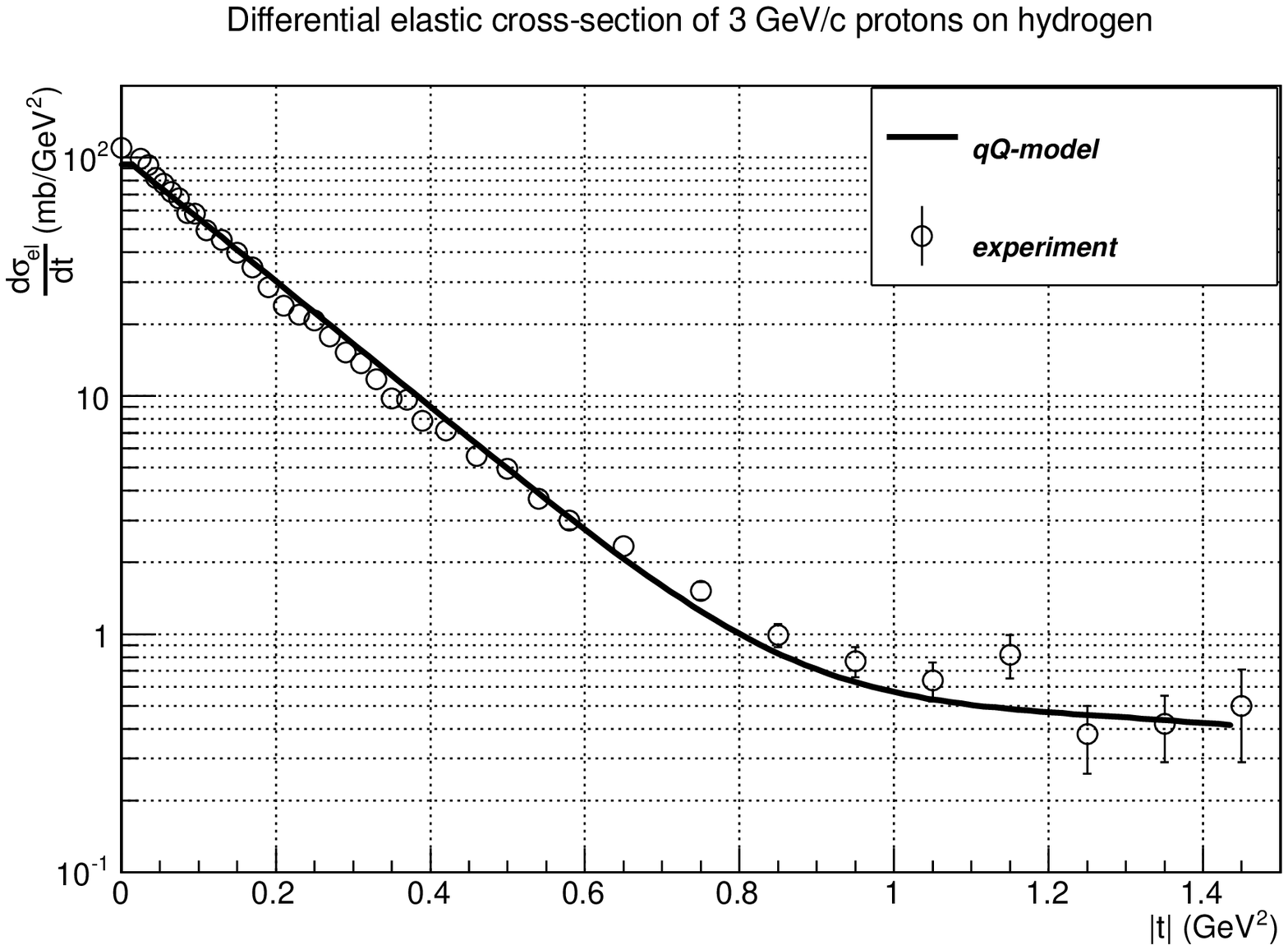}
\caption{The proton-proton differential elastic cross-section 
versus $|t|$ at the proton momentum in the laboratory system 3~GeV/c. The curve is the prediction of our model. 
The open circles are the experimental data~\cite{pp3gevc}. }
\label{pp3}
\end{figure}


\section{Summary}

We have considered the $qQ$-model of the nucleon-nucleon elastic scattering with springy Pomeron 
and compared it with experimental data. It was obtained reasonable description of  the 
differential cross section of elastic $pp$  $\bar{p}p$  and $np$ scattering in a wide range of 
energies from few GeV to 7~TeV in the center of mass system. The dip position and its value of 
the $d\sigma_{el}/dt$ are in the  satisfactory agreement with experimental data. 

The qQ-model, in the framework of relations (\ref{f1})-(\ref{f3}),  can be easily generalized to 
the case of elastic scattering of mesons on nucleons or mesons on mesons. Fig. 10 shows the 
$\pi^{+}-p$ differential elastic cross-section versus $|t|$ at the $\pi^{+}$ momentum in the 
laboratory system 4.122~GeV/c, as an example of the qQ-model application to the case of meson-nucleon 
elastic scattering. The meson-nucleon elastic scattering  in terms of the qQ-model with springy 
Pomeron will be reported in next paper.

\begin{figure}
\includegraphics[height=2.8in,width=3.5in]{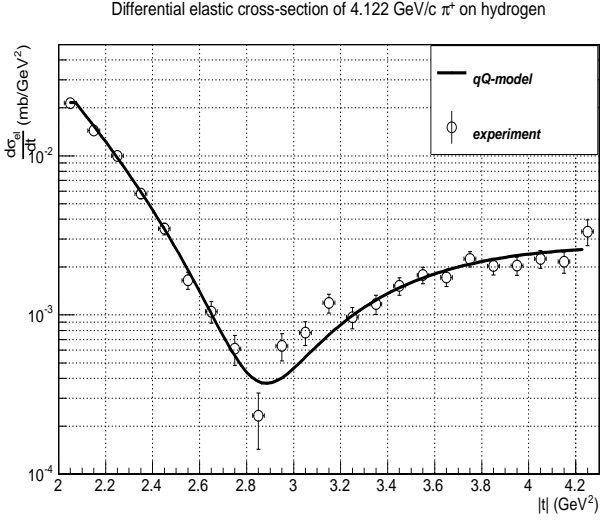}
\caption{The $\pi^{+}-p$ differential elastic cross-section 
versus $|t|$ at the $\pi^{+}$ momentum in the laboratory system 4.122~GeV/c. The curve is the prediction of our model. 
The open circles are the experimental data~\cite{pipp4gevc}. }
\label{pp3}
\end{figure}

\section*{Acknowledgment}
 
The author is thankful to S. Bertolucci, S. Giani and M. Mangano for stimulating discussions and support. 
The meetings and e-mail communication with N. Starkov and N. Zotov were powerful for clarification of 
the qQ-model details. The work was also  partly supported by the CERN-RAS Program of Fundamental 
Research at LHC.


\end{document}